# Farm Environmental Data Analyzer using a Decentralised system and R

Aryan Bagade
Department of Electronics and Telecommunication
Pune Institute of Computer Technology
Pune, India
aryanbagade@icloud.com

Prof. Rupesh C. Jaiswal
Department of Electronics and Telecommunication
Pune Institute of Computer Technology
Pune, India
rcjaiswal@pict.edu

*Abstract-* Data/Web Hosting is a service that lets enterprises or selves present their data on the internet that users can access. The firm providing such services are web/data host. Apart from that, such services require incessant support, and not everyone can afford a particular centralized data host service. The peer-to-peer(P2P) protocol, the Interplanetary file system(IPFS), is augmenting into a legitimate alternative to traditional data and web hosting. This paper put forward a decentralized blockchain IPFS-based interactive manageable model, and the work presents an application and schematic that serves as a Proof of Concept(PoC) of decentralized blockchain technology that can be wielded to create an immutable record of energy and resource usage for further analysis and estimation of yield at a scale. IPFS hosts an immutable record that would be independently verifiable and available in perpetuity. First, having the user connect to the service through an IPFS node, then requesting the user upload their data. Then the data is uploaded, and a CID(Content Identification) and a QR code are returned to the user, who can then compute and visualize the results through the application. This system enables the novel application of decentralized data storage to capture, add and visualize yield and environmental data and track it further down the supply chain.

## I. Introduction

People have witnessed the expeditious evolution of technology like mobile communication, the Internet of Things(IoT), high computational processors, etc. These technological advances have brought us to an era of big data. Every technology user generates enormous amounts of data all the time, and this ceaselessly generated large-scale data converges to multiple services like data hosts, data collectors, etc., which may tend to cause more and more identity and data leaks. Blockchain isn't mentioned or acknowledged in many technical systems for privacy safety. Nonetheless, blockchain is born with two essential attributes: trusted peers and immutable records, which are resistant to manipulation, which makes it distinct from the conventional solutions for privacy safety. Blockchain is about peer-to-peer communication on a decentralized network; in this decentralized system, peers can communicate directly, irrespective of location. The functions of the intermediaries are transferred to the periphery of the peer participants in the blockchain model.

The problem with the existing proof of work consensus is that data that can be stored on a block has a limited(that does not suit extensive data) size. To overcome this, the most promising solution that can be considered is the IPFS(Inter-Planetary File System). IPFS is a decentralized system for data transfer in contrast to centralized protocols, namespaces, and transfers provided by HTTP. It is about data distribution, content-based identification using a secure hash of data as a file location identifier, and resolving locations using Distributed Hash Table. The IPFS hash of the data stored by each node across the blocks is the same, which enables consistency simultaneously; this technique eliminates the constraint of dependence on the various full-node in the network.

The crux of this paper is to present a Farm Environmental Data Tracker and Analyser model based on a Decentralised system. The application is hosted on an IPFS network with IPFS RPC API address mentioned. RPC API is used for retrieving content-addressed data from the network. Then the farm data can be uploaded to the system, which will store it in the IPFS decentralized environment. Once the data is uploaded, the Hash of data or Content Identification (CID) is returned along with Data Visualiser Link and QR code to track the yield further in the supply chain. At this point, an immutable data record has been created that would be independently verifiable and further assessed. Using the Visualiser Link or in Visualise Farm Data window, Farm Environmental Data is statistically analyzed and viewed as graphs. This graph can draw conclusions and relations between different parameters like electricity, water, and fertilizer (expressed according to amount) versus the farm's yield. Additionally, the type of yield and location of the farm is also considered.

## II. Literature Survey

Blockchain is an immutable, Cryptographic, Peer-to-peer communication model. The globally widely used system put forth this technology is Bitcoin, first mentioned in the Bitcoin white paper [1] published by Nakamoto. The fundamental concept of the bitcoin network is Unspent Transaction Output(UTXO), a set of all UTXOs in a bitcoin system that defines the state of the Bitcoin blockchain. However, the expanding size of the UTXO set is downgrading the access performance [2] and drastically bringing down the validation speed of blockchain, particularly in resource-constrained applications like IoT [4].

In the Blockchain model, every node keeps a copy of the ledger, thus improving the security, but it has drawbacks. The trite of the same data leads to data redundancy. A network Coding-based distributed storage(NC-DS) framework [3] was formulated. This system was based on the block division method in which blocks are divided into several sub-blocks, and then using network encoding, these sub-blocks are further divided into more sub-blocks. The complete encoding and decoding process drastically increases the complexity as well as the efficiency of the system. There can be a possibility of data inconsistency problems.

Another paper [5] proposed an augmented summary block for the bitcoin blockchain. All UTXOs in the current block are documented in a file at regular intervals, and the remaining data is eliminated. Thus creating the drawback of not being able to retrace the history of the blockchain.

The mentioned schemes have several limitations, plus a hosting data service still needs to be solved. For this paper on the Farm Analyser system, IPFS(Inter-Planetary File System) is considered [6]. It's a peer-to-peer approach in which each node stores a compilation of hashed files. A client who wants to retrieve stored data must invoke the hash of the information it wants. IPFS then combs through the nodes and supplies the client with the data. One can think of it as similar to BitTorrent [9]. IPFS is content addressed. Hence, the same data can result in the same hash. The above scheme solves the problem of data inconsistency and can easily retain the traceability of system history. The synchronization speed between the newly added node and the network in the system is quickened [7]. The paper [8] proved that the storage compression ratio using IPFS is very effective and demonstrated on bitcoin blockchain data.

As far as Farm Environmental Data Analysis is concerned, the system uses R programming language [10]. It's one of the best and most widely used data reconfiguration and statistical analysis tools. Data wrangling, Data visualization, and research make it most prominent in the market. With the help of a shiny package in R [11], one can implement the whole application on the web.

**Motivation:-** The work in this project is motivated by aiming to make a decentralized hosted Analyser plus Visualiser system for the farm environmental data. There are various IoT-based innovative farm models [14, 15] through which data can be retrieved or accessed. Based on that data, we can configure it further across the IPFS network to create its verifiable, immutable record and derive viable results after the computation. Farm data can be monitored and analyzed, visualized, and automated using the system. Even help in tracking the yield for the further supply chain [13].

## III. IPFS TERMINOLOGIES

Let's start with an overview of the kernel block of our system, which is IPFS, and the fundamental principles of IPFS, which will be beneficial to the Farm Analyser system (i) Content Addressing; (ii) Peer Addressing; (iii) Content Indexing.

### A. Content Addressing

Content Addressing uses hash-based Content Identifiers (CID), analogous to BitTorrent, which is based on Content-Centric Networking [9]. CID is a self-defined content-addressed identifier, which doesn't indicate the location of data where it's stored, else it forms an address based on the content itself. The number of characters in CID doesn't depend upon the content size; rather, it depends upon the Cryptographic Hash. For example, if we stored our farm data on the IPFS network, its CID would look like this:

$Qmc9jWAvivgWrUdPoPEp7fo7sqRTaVEQr7Yb2NCcN7JAyJ$.

### B. Peer Addressing

One of the significant characteristics of a distributed network is the ability to access data by its content rather than its location [6]. It enables users to get data from any peer with the needed data. But what if a peer's location has been changed or transferred among multiple devices? For that, inside an IPFS network, peers specify each other via their Peer ID. For connecting with an exemplary peer, this ID provides a unique identity to each one of them. This unique identification and dealing with the correct peer is Peer Addressing. IPFS relies on Multiaddress to represent the location of remote peers. The Multiaddress is a self-defining, human-readable structure with an interconnected network and transport protocols for communication followed by PeerID.

### C. Content Indexing

It is mandatory to have a mapping between a CID and a Peer ID to upload or retrieve the data. These mappings are indexed on Distributed Hash Table (DHT), Aiming to make the system decentralized. IPFS's DHT is based on Kademlia [12], analogous to the Mainline DHT of BitTorrent [9]. Peers join the DHT either as DHT Servers or as DHT Clients depending on their connectivity, and IPFS distinguish between these through a procedure called Autonet [13]. By default, a new peer joins as a client and initiates the connection among other peers. Suppose more than three peers connect with the newly joined peer; then that new peer upgrades to DHT Server. DHT Client can only request the data but cannot store it, whereas the DHT server can store and provide it.

In Fig. 1, the critical terminologies on which IPFS is inspired are shown, and their relation to respective stack elements is illustrated. Stages of the data can also be derived from the stack; for example, the Data structure used in IPFS is based upon the data structure of Git, which is Merkle Data Acyclic Graph (DAG) [6], where the data is defined/stored.

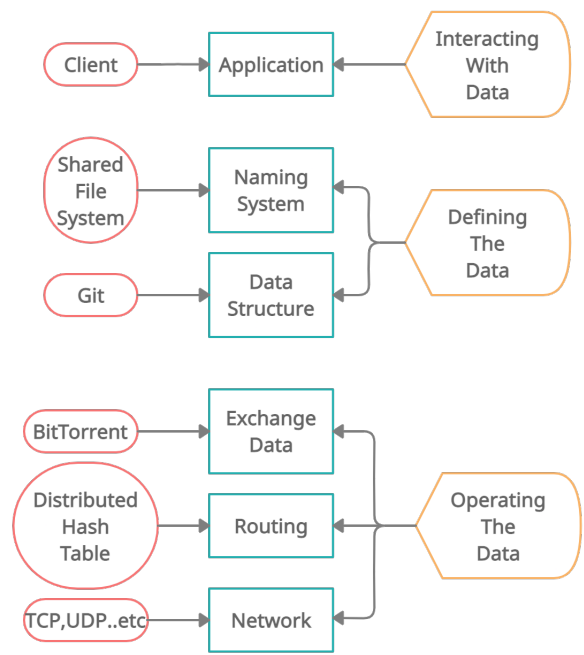

Figure 1: IPFS Stack with mentioned Terminologies

## IV. Implementation

This section describes the implementation of an innovative decentralized farm analyzer and its critical technologies. This proposed paper includes five effective workflows.

### A. IPFS Content Hosting

The critical difference between InterPlanetary File System (IPFS) and Hypertext Transfer Protocol (HTTP) is how the content is referenced. IPFS is content referenced, whereas HTTP is location referenced. The maximum benefit we get here is that one can retrieve the data from any peer who hosted it rather than retrieving it from the central server. When asking for the content on the internet using HTTP, we specify the range using the URL, but in the IPFS system, we access the content via Content Identifier (CID). Since the IPFS references Content, if the content alters, then CID will change. A CID is analogous to a URL.

The challenge with IPFS-hosted content is that most browsers cannot resolve an IPFS address, thus creating browser dependence. Brave and Opera browsers have native IPFS support. There are ways to access IPFS-hosted content on IPFS-avoided Browsers using IPFS Companion or being dependent on a Gateway [6]. A gateway is a middleware between the browser and IPFS. In this paper, to make an application browser-independent, we made an HTTP request to the server that fetches IPFS content. The only drawback is that the gateway is centralized and thus prone to shut down, but since the content is maintained on the IPFS, the content persists even after the gateway failure.

Gateway resolved the browser-dependent issue, but when one requests the content on IPFS, the receiving node also serves as a source for that content, and the garbage collector will discard the content after approximately 12 hours. For a robust application, eliminating data inside it is a severe issue. To avoid it, there's a concept known as Pinning. When a node pins the content on an IPFS, it'll be preserved forever (unless the user unpins it). Several pinning services can be configured with IPFS. Pinata is one of them and is used in the following application. For Pinata to work with the IPFS-based application, we need Pinata API. It's the main interface and is accountable for all the IPFS pinning, Uploading Content, Authentication, and more. To connect to it, we need Pinata API keys, as shown in Fig. 3; when you generate your keys, you get the set of Pinata API keys, Pinata API secret, and Pinata API JWT (JSON web token) Key [7]. This adds additional user security.

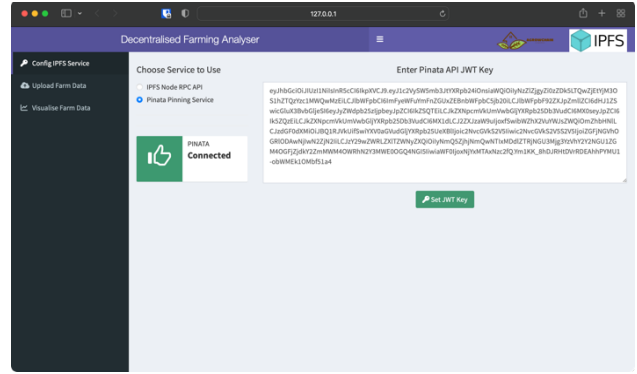

Figure 3: Pinata Service connected via valid JWT Key

The JWT key in Fig. 3 is an encrypted mixture of the Pinata API key and Pinata Secret API key. Since IPFS deals with public key (asymmetric key) cryptography [8], Pinata API Key acts as a public key, whereas the Pinata Secret API key acts as a private key.

When dealing with the local environment, initiating the IPFS daemon is mandatory [6].

### B. Data Uploading and Verification

Once the IPFS service is configured and hosting is done through the gateway, it is very straightforward to upload the data to an IPFS system and create its unalterable record, which can easily be verified. After retrieving the environmental data, farmers can easily upload it to an IPFS network through the application. Once the data is successfully uploaded, they'll receive CID, Visualiser Link, and QR code, as shown in Fig.5.

- CID:- Content Identifier to fetch the data from the IPFS environment.
- Visualizer Link:- To successfully analyze and evaluate the farm data and derive understandable graphs
- QR code:- Enables verifiable medium across the supply chain of the farm yield and promptly redirects the connection with the visualizer.

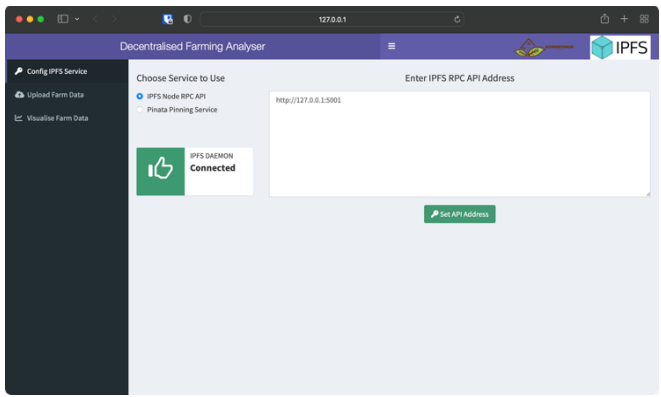

Figure 2: IPFS local node service using RPC API

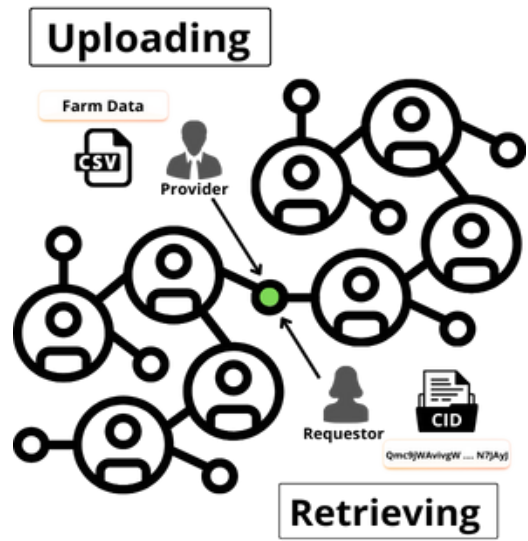

Figure 4: IPFS DHT Network transaction process

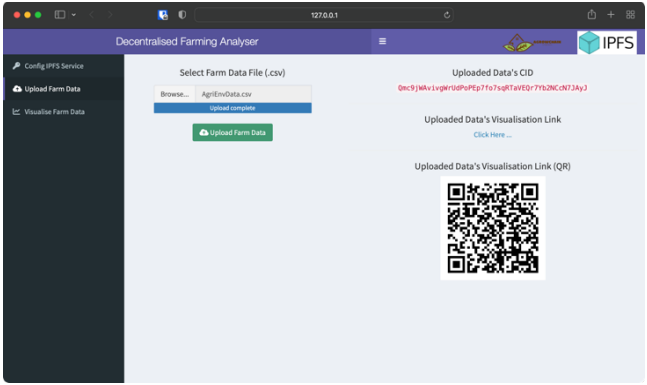

Figure 5: Application window to upload the Farm data

In Fig. 4, the IPFS transaction process. **Uploading involves** me) Uploading the farm data to the local IPFS node and retrieving the CID. ii) DHT plays its role in discovering the closest peer to CID. iii) Uploader record with the closest peer. **Retrieval involves** iv) request of Bitswap to already bounded peers for CID. v) DHT plays its role in finding a farm data-storing peer. vi) Requestor connects to uploader/provider and fetches the farm data. Then Uploader's PeerID is mapped to their respective network address [12].

*C. Data Evaluation*

As the Farm data is published on an IPFS, anyone with the respective CID can access and verify the data. In the third workflow, farm data will be evaluated and presented as graphs. From these results, we can derive the efficient production of yield in different use cases. Farm Visualiser fetches the Farm data from the IPFS, evaluates it with various R queries in the backend, and presents it on an application window. R has the package for almost every kind of representation. One can even redirect to a visualizer using the Visualiser link or the QR code presented in the Upload Farm Data section, Fig. 5.

How will the application retrieve data and compute it? In the previous section, Peer and Uploader records have been published, and almost anyone with a valid CID can retrieve the content. To fetch the data, the application performs four steps i)Data Finding using CID and its correlated PeerID; ii) Finding Data Provider; iii) Routing or Connection; iv)Data Exchange. All the terminologies are part of the IPFS stack mentioned in Fig. 1

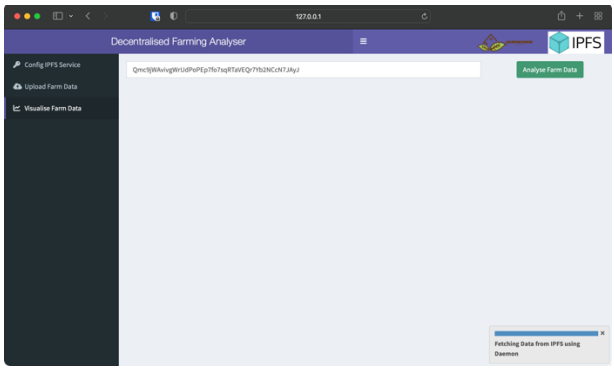

Figure 6: Farm data Visualizer Window

In Fig. 6, we can see that based on the Content Identifier, the application starts fetching Farm data from IPFS using the daemon. It's to be noted that whenever a user retrieves that data, that respective user becomes a Data provider themselves by uploading the provider contract to their node to Distributed Hash Table.

*D. Codebase*

As far as the Analyzer's codebase is concerned, the whole system is developed in the R language. R is undoubtedly known for its great modeling and development capabilities. DT, shiny and shinydashboard packages of R have been used to develop web applications [11]. tidyverse and lubridate packages of R have been used to model the graphical representation of Data [10]. The file structure of the system consists of three R files (ui.R, server.R, ipfs.R), one txt file (jwt_key.txt), and an asset folder (www) as shown in Fig. 7. To run the system on a local environment start the IPFS daemon and then logs in R panel and enter a query

$$shiny :: runApp()$$

The complete system's codebase is open-sourced and viewed on GitHub [16]

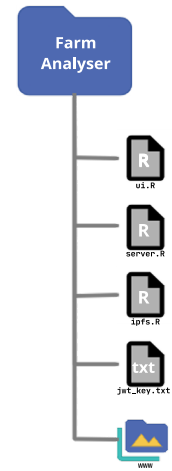

Figure 7: Farm Analyzer System file structure

V. RESULTS AND DEMONSTRATION

In the previous sections, we have set up the environment, hosted the content, uploaded the Farm data, Retrieved it from the IPFS network, and Evaluated it using R packages and queries. Now it's time to see the evaluated results. The results will be represented in graphs in which Yield will be compared with parameters like Total time, Water Consumption, Fertiliser, and Electricity. Additionally, one can even select a particular product type and farm location into consideration.

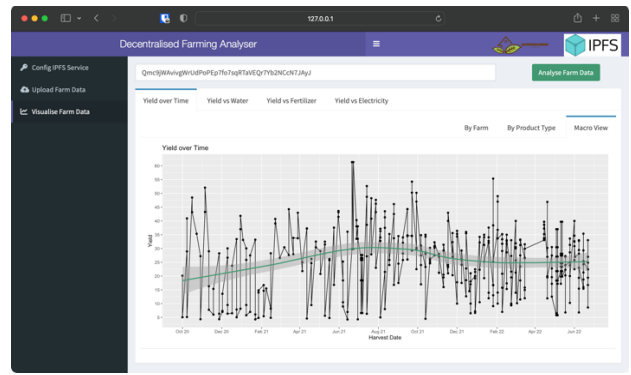

Figure 8: Yield over time in Macro View

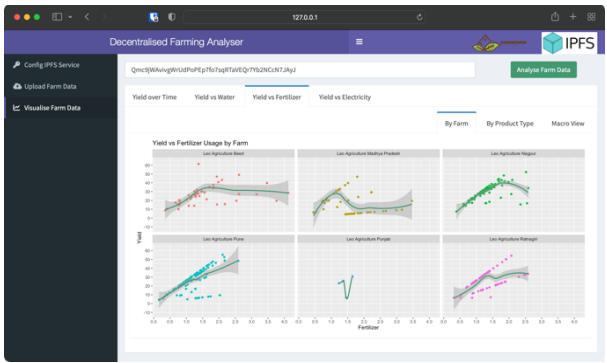

Figure 9: Yield vs. Fertilizers by Farm type

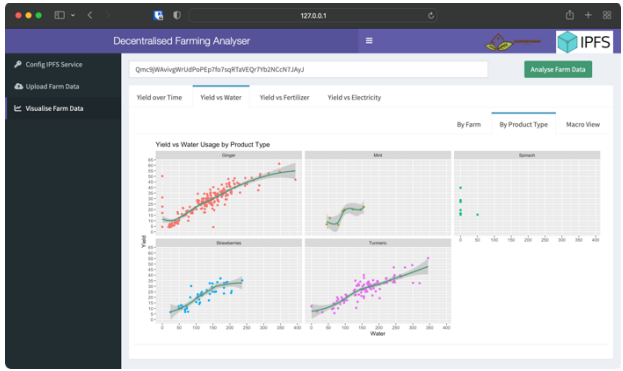

Figure 10: Yield vs. Water by Product type

After the demonstration, the Analysis of the total bandwidth required and consumption of the resources are presented below with the format in which our data was saved on an IPFS network.

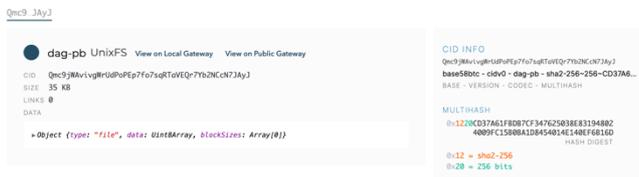

Figure 11: CID and Multiaddress of the content uploaded

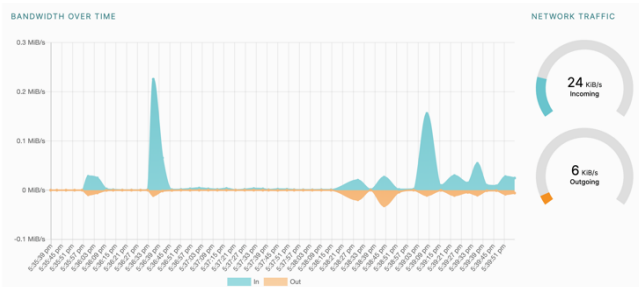

Figure 12: Bandwidth required over time

## VI. CONCLUSION AND FUTURE SCOPE

This study aims to develop a Decentralised Farm Environmental data analyzer. This system has been implemented by considering many crucial parameters like privacy, scalability, and primarily distributed hosting of data. The plan was made with the aim that it can be used by anyone who has minimal technical proficiency. The proposed system will take farm data from IoT-based conventional or vertical farms, Upload it to the decentralized IPFS network to create its verifiable, immutable record, and evaluate it to present the visual relational graphs. Additionally, support the yield producer to the consumer supply chain.

This work can be further extended by integrating it with a decentralized mobile application with a machine learning model to recommend the most viable yield concerning the location, resources, environment, and much more.

## References

[1] Nakamoto, Satoshi. (2009). Bitcoin: A Peer-to-Peer Electronic Cash System. Cryptography Mailing list at https://metzdowd.com.

[2] S. Jiang et al., "BZIP: A Compact Data Memory System for UTXO-based Blockchains," 2019 IEEE International Conference on Embedded Software and Systems (ICESS), 2019, pp. 1-8, doi: 10.1109/ICESS.2019.8782459.

[3] A. Palai, M. Vora and A. Shah, "Empowering Light Nodes in Blockchains with Block Summarization," 2018 9th IFIP International Conference on New Technologies, Mobility and Security (NTMS), 2018, pp. 1-5, doi: 10.1109/NTMS.2018.8328735.

[4] A. Dorri, S. S. Kanhere, R. Jurdak and P. Gauravaram, "Blockchain for IoT security and privacy: The case study of a smart home," 2017 IEEE International Conference on Pervasive Computing and Communications Workshops (PerCom Workshops), 2017, pp. 618-623, doi: 10.1109/PERCOMW.2017.7917634.

[5] Gao, J., Li, B., Li, Z. (2020). Blockchain Storage Analysis and Optimization of Bitcoin Miner Node. In: Liang, Q., Liu, X., Na, Z., Wang, W., Mu, J., Zhang, B. (eds) Communications, Signal Processing, and Systems. CSPS 2018. Lecture Notes in Electrical Engineering, vol 517. Springer, Singapore. https://doi.org/10.1007/978-981-13-6508-9_112

[6] Benet, Juan. (2014). IPFS - Content Addressed, Versioned, P2P File System

[7] Trautwein, D., Raman, A., Tyson, G., Castro, I., Scott, W., Schubotz, M., Gipp, B. and Psaras, Y., 2022. Design and evaluation of IPFS: a storage layer for the decentralized web. arXiv preprint arXiv:2208.05877

[8] Q. Zheng, Y. Li, P. Chen and X. Dong, "An Innovative IPFS-Based Storage Model for Blockchain," 2018 IEEE/WIC/ACM International Conference on Web Intelligence (WI), 2018, pp. 704-708, doi: 10.1109/WI.2018.000-8.

[9] Cohen, Bram. "The bittorrent protocol specification." (2008).

[10] Petzoldt, Thomas. (2017). Data Analysis with R: Selected Topics and Examples.

[11] Lihua Jia, Wen Yao, Yingru Jiang, Yang Li, Zhizhan Wang, Haoran Li, Fangfang Huang, Jiaming Li, Tiantian Chen, Huiyong Zhang, Development of interactive biological web applications with R/Shiny, *Briefings in Bioinformatics*, Volume 23, Issue 1, January 2022, bbab415, https://doi.org/10.1093/bib/bbab415

[12] Wang, Liang, and Jussi Kangasharju. "Measuring large-scale distributed systems: case of bittorrent mainline dht." *IEEE P2P 2013 Proceedings*. IEEE, 2013.

[13] Protocol Labs. 2022. AutoNAT. Retrieved 01 June 2022 from https://github.com/ libp2p/specs/blob/master/autonat/README.md

[14] C. Yoon, M. Huh, S. -G. Kang, J. Park and C. Lee, "Implement smart farm with IoT technology," 2018 20th International Conference on Advanced Communication Technology (ICACT), 2018, pp. 749-752, doi: 10.23919/ICACT.2018.8323908.

[15] O. Chieochan, A. Saokaew and E. Boonchieng, "IOT for smart farm: A case study of the Lingzhi mushroom farm at Maejo University," 2017 14th International Joint Conference on Computer Science and Software Engineering (JCSSE), 2017, pp. 1-6, doi: 10.1109/JCSSE.2017.8025904.

[16] Bagade (2022) Farm Data Analyser [Source Code]. https://github.com/AryanBagade/FarmingAnalyzer.